\documentstyle[11pt,adassconf]{article}  

\begin{document}   

%
%
%

\paperID{P1-25}

%
%

\title{Automated spectral extraction for high multiplexing MOS and IFU
	observations}

%
%
%

\author{Marco Scodeggio\altaffilmark{1}, Alessandra Zanichelli, Bianca Garilli}
\affil{IFC--CNR, Milano, Italy}
\author{Olivier Le F{\`e}vre}
\affil{LAM--CNRS, Marseille, France}
\author{Giampaolo Vettolani}
\affil{IRA--CNR, Bologna, Italy}
\altaffiltext{1}{On behalf of the VIRMOS Consortium}
%
%

\contact{Marco Scodeggio}
\email{marcos@ifctr.mi.cnr.it}

%
%
%

\paindex{Scodeggio, M.}
\aindex{Zanichelli, A.}     
\aindex{Garilli, B.}
\aindex{Le F{\`e}vre, O.}
\aindex{Vettolani, G.}
%
%



\begin{abstract}          
VIMOS main distinguishing characteristic is its very high multiplex
capability: in MOS mode up to 800 spectra can be acquired
simultaneously, while the Integral Field Unit produces 6400 spectra to
obtain integral field spectroscopy of an area approximately 1$\times$1
arcmin in size.  To successfully exploit the capabilities of such an
instrument it is necessary to expedite as much as possible the
analysis of the very large volume of data that it will produce,
automating almost completely the basic data reduction and
the related bookkeeping process.
The VIMOS Data Reduction Software (DRS) has been designed specifically
to satisfy these two requirements. A complete automation is achieved
using a series of auxiliary tables that store all the input
information needed by the data reduction procedures, and all the
output information that they produce.  We expect to achieve a
satisfactory data reduction for more than 90\% of the input spectra,
while some level of human intervention might be required for a small
fraction of them to complete the data reduction.
The DRS procedures can be used as a stand-alone package, but are also
being incorporated within the VIMOS pipeline under development at the
European Southern Observatory.  

\end{abstract}

%
%

\section{Introduction}

VIMOS is the first of a pair of imaging spectrographs that are being
built for the unit telescopes of the European Southern Observatory
Very Large Telescope. The instrument field of view is split into four
separate quadrants, each one covering approximately $7\times8$
arcmin. On one side of the instrument is anchored the head of the
Integral Field Unit (IFU), consisting of a lenslet array of 6400
lenslets organized in an $80\times80$ array that covers a field of
view of $54\times54$ arcsec. The light collected by the IFU head is
fed to the main spectrograph via optical fibers.

This instrument was designed specifically to carry out survey work,
and to have a very high multiplexing capability. Because of this, it
will also require a rather new approach to the process of data
reduction: it will be simply impossible to reduce by hand the very
large amount of data that VIMOS will produce, as the following example
demonstrates. Working ``by hand'' using IRAF (or a similar package),
even with a number of {\it ad hoc} scripts designed to carry out
repetitive tasks, an astronomer could reduce one spectrum in 3 to 5
minutes. As VIMOS in Multi Object Spectrograph (MOS) mode will produce
between 150 and 200 spectra per quadrant for each exposure, and it has
4 quadrants, it would then take between 30 and 70 hours of work to
reduce a single exposure. Therefore, to reduce just one night of
observations ``by hand'' it would require between 300 and 700 hours of
work, which is 40 to 150 working days (8 hours a day, doing nothing
else)!! With IFU observations, each one producing 6400 spectra, things
would obviously be even worse.

For this reason it was decided to implement a completely automatic
data reduction pipeline for the processing of VIMOS data. As part of
the agreement for the construction of the instrument, the VIRMOS
consortium is developing the core components of this pipeline, what we
call here the VIMOS Data Reduction Software (DRS). The European
Southern Observatory is in charge of putting this core component into
a completely automatic pipeline within the framework of their Data
Flow System. Although the DRS has been designed specifically for
VIMOS, its conceptual lay-out is general enough that it could be
easily adapted to any kind of instrument with capabilities comparable
to those of VIMOS.

\section{Instrument and individual mask calibrations}

It is very difficult to design completely automated tasks to work
blindly at the reduction of complex datasets like those produced by
VIMOS. It was thus decided that the DRS will {\it always and only}
work starting from a ``reasonable'' first guess about all calibration
parameters of the instrument. This will require the setup of an
instrument calibration database, and the periodical execution of
calibration procedures that will produce the necessary calibration
data. When reduced using DRS procedures, these data will produce:

\noindent\hangindent 15pt 1) a mapping between positions on the
slitlets mask and positions on the CCD frame where the slitlets images
are recorded;

\noindent\hangindent 15pt 2) a mapping of the distortions that affect
each individual spectrum;

\noindent\hangindent 15pt 3) a mapping of the wavelength dispersion
solution, that associates a wavelength to each pixel coordinate of a
spectrum image;

\noindent\hangindent 15pt 4) two mappings between celestial
coordinates and coordinates in the plane of the slitlets mask and in
the plane of the CCD frame.

\noindent
These mappings, generally in the form of second to fourth order
spatial polynomials, are stored in calibration matrices coefficients
inside FITS binary tables.

\begin{figure}
\epsscale{0.5}
\plotone{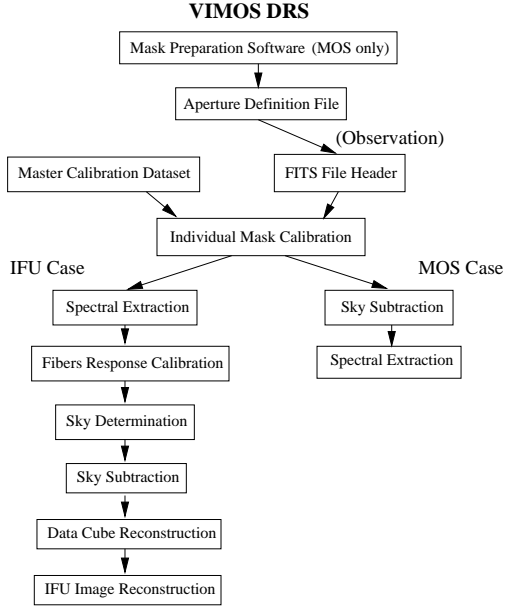}
\caption{A block diagram scheme of the DRS spectral extraction process}
\end{figure}


Starting from the general instrument calibrations, specific
calibrations are obtained for each specific mask used for an
observation (see Figure 1):

\noindent\hangindent 15pt 1) the Mask Preparation Software used by the
astronomer to define the slitlets positions for MOS observations (see
Bottini, Garilli \& Tresse 2000) produces an Aperture Definition File
(ADF), containing the positions of all slitlets in the mask plane;

\noindent\hangindent 15pt 2) the ADF is inserted into the FITS file
header at the end of the observation (for IFU observations, where the
ADF is always the same, this is stored in a separate table);

\noindent\hangindent 15pt 3) using the slit positions derived from the
ADF, and the general calibration matrices, the approximate location of
each two-dimensional spectrum on the CCD frame is computed;

\noindent\hangindent 15pt 4) a flat field exposure is used to measure
the exact location of each two-dimensional spectrum on the CCD frame;

\noindent\hangindent 15pt 5) a calibration lamp exposure is used to
measure the position of a number of spectral lines and refine the
approximate wavelength solution provided by the general calibration matrix.

\section{Spectral Extraction}

The final result of the calibration phase is the complete definition
of a set of apertures to be used for the final spectral
extraction. The extraction itself is carried out following two
different procedures for MOS and IFU observations.

For observations carried out in MOS mode, it is assumed that slitlets
always contain a region of ``pure sky'', where the contribution of the
sky background to the composite spectrum of sky plus target
astronomical object can be estimated. Each two-dimensional spectrum is
collapsed along the wavelength dispersion axis, to produce a
mono-dimensional intensity profile along the entrance slit. Groups of
at least 3 adjacent pixels all with intensity above a certain
threshold (derived from the rms noise in the profile itself) are
considered as ``object'', while all other pixels are considered as
``sky''. One-dimensional spectra for each object so identified are
obtained using Horne (1986) optimal extraction method.

For observations carried out in IFU mode, first of all one-dimensional
spectra for each lenslet are obtained adding together the flux
collected by the corresponding optical fiber.  Each one of these
spectra can belong to one of two categories: either it is the
superposition of sky and astronomical object contributions, or it is a
pure sky spectrum. It is therefore necessary to identify those spectra
in this second category to build an estimate of the sky background
intensity, before subtracting it from all the spectra. This
identification is done on the basis of the distribution of total light
intensities registered in the various spectra, considering as pure sky
ones those that have an intensity lower than that corresponding to the
mode of the distribution.  For extremely crowded fields (or very deep
exposures), the number of pure sky spectra is expected to be a very
small fraction of the total number of spectra. In this case an
interactive tool will be provided, to allow the astronomer to specify
``by hand'' the spectra to be used to build the sky background
estimate.

After the pure sky spectra are identified, they are median-averaged
together, to build an estimate of the sky spectrum, which is
subtracted from all the 6400 one-dimensional spectra.  Since the
optical fibers redistribute the light collected by the lenslet array
over the four VIMOS quadrants following a rather complex pattern, the
one-dimensional sky-subtracted spectra are then re-assembled in a data
cube to provide a complete and spatially coherent reconstruction of
the given data-set. Also, all spectra can be collapsed along the
wavelength axis to provide a two-dimensional image of the area covered
by the IFU field of view.

\section{Final Remarks}

The VIMOS DRS is written entirely using ANSI C code. Highly specialized
tasks are carried out by external libraries, including the CFITSIO and
WCS libraries, and the SExtractor object detection software.
All input and output files are standard FITS files (image and
binary table format). Heavy usage is made of ESO hierarchical keywords
within headers (this means IRAF would have problems handling those
headers). 

The final DRS output will be: for MOS observations a 2-dimensional
FITS image, containing one extracted spectrum for each row of the
image, plus a table containing the identification parameters for each
one of those spectra; for IFU observations a 3-dimensional FITS image
containing the IFU data cube, plus a table containing the
identification parameters for each spectrum.


\begin{references}

\reference Bottini, D., Garilli, B., \& Tresse, L.\ 2000, \adassx, 
	\paperref{P1-26}
\reference Horne, K. 1986, \pasp, 98, 609

\end{references}
\end{document}